\documentstyle[12pt]{article}
\begin{document}

\begin{flushright}
ITKP 95-22\\
ISN 95-87\\
\today
\end{flushright}
\vskip 1cm
\begin{center}
\baselineskip=.5cm
{\Large\bf Remarks about spin measurements}
\vskip .2cm
{\Large\bf in
\boldmath$\bar{\rm p}{\rm p}  \rightarrow  
\overline{\Lambda}{\Lambda}$
\unboldmath}
\vskip .5cm
{Jean-Marc Richard\footnote{e-mail: {\tt jmrichar@isnnx2.in2p3.fr}}}
\vskip .2cm
{\small Institut des Sciences Nucl\'eaires}\\
{\small Universit\'e Joseph Fourier--CNRS-IN2P3}\\
{\small 53, avenue des Martyrs, 38026 Grenoble Cedex, France}\\
{\small and}\\
{\small Institut f{\"u}r Theoretische Kernphysik}\\
{\small Rheinische Friedrich-Wilhelms Universit{\"a}t}\\
{\small Nu\ss allee 14-16, 53115 Bonn, Germany}

\vskip .5cm
{\small\bf Abstract}

\vskip .2cm
\begin{minipage}{120mm}
\baselineskip=.5cm
{\small
  We comment on the proposed measurement of the spin-depolarisation
  parameter $D_{nn}$ in the strangeness-exchange reaction $\bar{\rm
    p}{\rm p} \rightarrow \overline{\Lambda}{\Lambda}$. It is shown  
that
  the existing data on the correlation coefficients $C_{ij}$ limit  
the range
  allowed for $D_{nn}$.}
\end{minipage}
\end{center}

\vskip .5cm

\begin{figure}[htbc]


\setbox1=\vbox{
\setlength{\unitlength}{0.240900pt}
\ifx\plotpoint\undefined\newsavebox{\plotpoint}\fi
\sbox{\plotpoint}{\rule[-0.200pt]{0.400pt}{0.400pt}}%
\begin{picture}(900,600)(0,0)
\font\gnuplot=cmr10 at 10pt
\gnuplot
\sbox{\plotpoint}{\rule[-0.200pt]{0.400pt}{0.400pt}}%
\put(220.0,323.0){\rule[-0.200pt]{148.394pt}{0.400pt}}
\put(220.0,170.0){\rule[-0.200pt]{4.818pt}{0.400pt}}
\put(198,170){\makebox(0,0)[r]{$-0.6$}}
\put(816.0,170.0){\rule[-0.200pt]{4.818pt}{0.400pt}}
\put(220.0,272.0){\rule[-0.200pt]{4.818pt}{0.400pt}}
\put(198,272){\makebox(0,0)[r]{$-0.2$}}
\put(816.0,272.0){\rule[-0.200pt]{4.818pt}{0.400pt}}
\put(220.0,373.0){\rule[-0.200pt]{4.818pt}{0.400pt}}
\put(198,373){\makebox(0,0)[r]{$0.2$}}
\put(816.0,373.0){\rule[-0.200pt]{4.818pt}{0.400pt}}
\put(220.0,475.0){\rule[-0.200pt]{4.818pt}{0.400pt}}
\put(198,475){\makebox(0,0)[r]{$0.6$}}
\put(816.0,475.0){\rule[-0.200pt]{4.818pt}{0.400pt}}
\put(220.0,577.0){\rule[-0.200pt]{4.818pt}{0.400pt}}
\put(198,577){\makebox(0,0)[r]{$1$}}
\put(816.0,577.0){\rule[-0.200pt]{4.818pt}{0.400pt}}
\put(220.0,68.0){\rule[-0.200pt]{0.400pt}{4.818pt}}
\put(220,23){\makebox(0,0){$-1$}}
\put(220.0,557.0){\rule[-0.200pt]{0.400pt}{4.818pt}}
\put(343.0,68.0){\rule[-0.200pt]{0.400pt}{4.818pt}}
\put(343,23){\makebox(0,0){$-0.6$}}
\put(343.0,557.0){\rule[-0.200pt]{0.400pt}{4.818pt}}
\put(466.0,68.0){\rule[-0.200pt]{0.400pt}{4.818pt}}
\put(466,23){\makebox(0,0){$-0.2$}}
\put(466.0,557.0){\rule[-0.200pt]{0.400pt}{4.818pt}}
\put(590.0,68.0){\rule[-0.200pt]{0.400pt}{4.818pt}}
\put(590,23){\makebox(0,0){$0.2$}}
\put(590.0,557.0){\rule[-0.200pt]{0.400pt}{4.818pt}}
\put(713.0,68.0){\rule[-0.200pt]{0.400pt}{4.818pt}}
\put(713,23){\makebox(0,0){$0.6$}}
\put(713.0,557.0){\rule[-0.200pt]{0.400pt}{4.818pt}}
\put(836.0,68.0){\rule[-0.200pt]{0.400pt}{4.818pt}}
\put(836,23){\makebox(0,0){$1$}}
\put(836.0,557.0){\rule[-0.200pt]{0.400pt}{4.818pt}}
\put(220.0,68.0){\rule[-0.200pt]{148.394pt}{0.400pt}}
\put(836.0,68.0){\rule[-0.200pt]{0.400pt}{122.618pt}}
\put(220.0,577.0){\rule[-0.200pt]{148.394pt}{0.400pt}}
\put(45,322){\makebox(0,0){$\; P$}}
\put(220.0,68.0){\rule[-0.200pt]{0.400pt}{122.618pt}}
\put(251,276){\raisebox{-.8pt}{\makebox(0,0){$\Diamond$}}}
\put(312,205){\raisebox{-.8pt}{\makebox(0,0){$\Diamond$}}}
\put(374,208){\raisebox{-.8pt}{\makebox(0,0){$\Diamond$}}}
\put(436,209){\raisebox{-.8pt}{\makebox(0,0){$\Diamond$}}}
\put(497,159){\raisebox{-.8pt}{\makebox(0,0){$\Diamond$}}}
\put(559,245){\raisebox{-.8pt}{\makebox(0,0){$\Diamond$}}}
\put(620,267){\raisebox{-.8pt}{\makebox(0,0){$\Diamond$}}}
\put(667,334){\raisebox{-.8pt}{\makebox(0,0){$\Diamond$}}}
\put(697,444){\raisebox{-.8pt}{\makebox(0,0){$\Diamond$}}}
\put(728,380){\raisebox{-.8pt}{\makebox(0,0){$\Diamond$}}}
\put(759,392){\raisebox{-.8pt}{\makebox(0,0){$\Diamond$}}}
\put(790,409){\raisebox{-.8pt}{\makebox(0,0){$\Diamond$}}}
\put(821,377){\raisebox{-.8pt}{\makebox(0,0){$\Diamond$}}}
\put(251.0,245.0){\rule[-0.200pt]{0.400pt}{14.936pt}}
\put(241.0,245.0){\rule[-0.200pt]{4.818pt}{0.400pt}}
\put(241.0,307.0){\rule[-0.200pt]{4.818pt}{0.400pt}}
\put(312.0,175.0){\rule[-0.200pt]{0.400pt}{14.454pt}}
\put(302.0,175.0){\rule[-0.200pt]{4.818pt}{0.400pt}}
\put(302.0,235.0){\rule[-0.200pt]{4.818pt}{0.400pt}}
\put(374.0,179.0){\rule[-0.200pt]{0.400pt}{14.213pt}}
\put(364.0,179.0){\rule[-0.200pt]{4.818pt}{0.400pt}}
\put(364.0,238.0){\rule[-0.200pt]{4.818pt}{0.400pt}}
\put(436.0,180.0){\rule[-0.200pt]{0.400pt}{13.972pt}}
\put(426.0,180.0){\rule[-0.200pt]{4.818pt}{0.400pt}}
\put(426.0,238.0){\rule[-0.200pt]{4.818pt}{0.400pt}}
\put(497.0,127.0){\rule[-0.200pt]{0.400pt}{15.418pt}}
\put(487.0,127.0){\rule[-0.200pt]{4.818pt}{0.400pt}}
\put(487.0,191.0){\rule[-0.200pt]{4.818pt}{0.400pt}}
\put(559.0,215.0){\rule[-0.200pt]{0.400pt}{14.213pt}}
\put(549.0,215.0){\rule[-0.200pt]{4.818pt}{0.400pt}}
\put(549.0,274.0){\rule[-0.200pt]{4.818pt}{0.400pt}}
\put(620.0,240.0){\rule[-0.200pt]{0.400pt}{13.249pt}}
\put(610.0,240.0){\rule[-0.200pt]{4.818pt}{0.400pt}}
\put(610.0,295.0){\rule[-0.200pt]{4.818pt}{0.400pt}}
\put(667.0,306.0){\rule[-0.200pt]{0.400pt}{13.490pt}}
\put(657.0,306.0){\rule[-0.200pt]{4.818pt}{0.400pt}}
\put(657.0,362.0){\rule[-0.200pt]{4.818pt}{0.400pt}}
\put(697.0,416.0){\rule[-0.200pt]{0.400pt}{13.249pt}}
\put(687.0,416.0){\rule[-0.200pt]{4.818pt}{0.400pt}}
\put(687.0,471.0){\rule[-0.200pt]{4.818pt}{0.400pt}}
\put(728.0,355.0){\rule[-0.200pt]{0.400pt}{12.045pt}}
\put(718.0,355.0){\rule[-0.200pt]{4.818pt}{0.400pt}}
\put(718.0,405.0){\rule[-0.200pt]{4.818pt}{0.400pt}}
\put(759.0,370.0){\rule[-0.200pt]{0.400pt}{10.600pt}}
\put(749.0,370.0){\rule[-0.200pt]{4.818pt}{0.400pt}}
\put(749.0,414.0){\rule[-0.200pt]{4.818pt}{0.400pt}}
\put(790.0,388.0){\rule[-0.200pt]{0.400pt}{9.877pt}}
\put(780.0,388.0){\rule[-0.200pt]{4.818pt}{0.400pt}}
\put(780.0,429.0){\rule[-0.200pt]{4.818pt}{0.400pt}}
\put(821.0,358.0){\rule[-0.200pt]{0.400pt}{9.154pt}}
\put(811.0,358.0){\rule[-0.200pt]{4.818pt}{0.400pt}}
\put(811.0,396.0){\rule[-0.200pt]{4.818pt}{0.400pt}}
\sbox{\plotpoint}{\rule[-0.500pt]{1.000pt}{1.000pt}}%
\put(251,288){\circle*{18}}
\put(312,241){\circle*{18}}
\put(374,196){\circle*{18}}
\put(436,244){\circle*{18}}
\put(497,203){\circle*{18}}
\put(559,212){\circle*{18}}
\put(620,241){\circle*{18}}
\put(667,286){\circle*{18}}
\put(697,337){\circle*{18}}
\put(728,321){\circle*{18}}
\put(759,335){\circle*{18}}
\put(790,353){\circle*{18}}
\put(821,343){\circle*{18}}
\multiput(251,267)(0.000,20.756){3}{\usebox{\plotpoint}}
\put(251,310){\usebox{\plotpoint}}
\put(241.00,267.00){\usebox{\plotpoint}}
\put(261,267){\usebox{\plotpoint}}
\put(241.00,310.00){\usebox{\plotpoint}}
\put(261,310){\usebox{\plotpoint}}
\multiput(312,219)(0.000,20.756){3}{\usebox{\plotpoint}}
\put(312,263){\usebox{\plotpoint}}
\put(302.00,219.00){\usebox{\plotpoint}}
\put(322,219){\usebox{\plotpoint}}
\put(302.00,263.00){\usebox{\plotpoint}}
\put(322,263){\usebox{\plotpoint}}
\multiput(374,177)(0.000,20.756){2}{\usebox{\plotpoint}}
\put(374,215){\usebox{\plotpoint}}
\put(364.00,177.00){\usebox{\plotpoint}}
\put(384,177){\usebox{\plotpoint}}
\put(364.00,215.00){\usebox{\plotpoint}}
\put(384,215){\usebox{\plotpoint}}
\multiput(436,226)(0.000,20.756){2}{\usebox{\plotpoint}}
\put(436,262){\usebox{\plotpoint}}
\put(426.00,226.00){\usebox{\plotpoint}}
\put(446,226){\usebox{\plotpoint}}
\put(426.00,262.00){\usebox{\plotpoint}}
\put(446,262){\usebox{\plotpoint}}
\multiput(497,185)(0.000,20.756){2}{\usebox{\plotpoint}}
\put(497,221){\usebox{\plotpoint}}
\put(487.00,185.00){\usebox{\plotpoint}}
\put(507,185){\usebox{\plotpoint}}
\put(487.00,221.00){\usebox{\plotpoint}}
\put(507,221){\usebox{\plotpoint}}
\multiput(559,194)(0.000,20.756){2}{\usebox{\plotpoint}}
\put(559,229){\usebox{\plotpoint}}
\put(549.00,194.00){\usebox{\plotpoint}}
\put(569,194){\usebox{\plotpoint}}
\put(549.00,229.00){\usebox{\plotpoint}}
\put(569,229){\usebox{\plotpoint}}
\multiput(620,225)(0.000,20.756){2}{\usebox{\plotpoint}}
\put(620,258){\usebox{\plotpoint}}
\put(610.00,225.00){\usebox{\plotpoint}}
\put(630,225){\usebox{\plotpoint}}
\put(610.00,258.00){\usebox{\plotpoint}}
\put(630,258){\usebox{\plotpoint}}
\multiput(667,265)(0.000,20.756){3}{\usebox{\plotpoint}}
\put(667,307){\usebox{\plotpoint}}
\put(657.00,265.00){\usebox{\plotpoint}}
\put(677,265){\usebox{\plotpoint}}
\put(657.00,307.00){\usebox{\plotpoint}}
\put(677,307){\usebox{\plotpoint}}
\multiput(697,319)(0.000,20.756){2}{\usebox{\plotpoint}}
\put(697,354){\usebox{\plotpoint}}
\put(687.00,319.00){\usebox{\plotpoint}}
\put(707,319){\usebox{\plotpoint}}
\put(687.00,354.00){\usebox{\plotpoint}}
\put(707,354){\usebox{\plotpoint}}
\multiput(728,306)(0.000,20.756){2}{\usebox{\plotpoint}}
\put(728,336){\usebox{\plotpoint}}
\put(718.00,306.00){\usebox{\plotpoint}}
\put(738,306){\usebox{\plotpoint}}
\put(718.00,336.00){\usebox{\plotpoint}}
\put(738,336){\usebox{\plotpoint}}
\multiput(759,321)(0.000,20.756){2}{\usebox{\plotpoint}}
\put(759,348){\usebox{\plotpoint}}
\put(749.00,321.00){\usebox{\plotpoint}}
\put(769,321){\usebox{\plotpoint}}
\put(749.00,348.00){\usebox{\plotpoint}}
\put(769,348){\usebox{\plotpoint}}
\multiput(790,341)(0.000,20.756){2}{\usebox{\plotpoint}}
\put(790,364){\usebox{\plotpoint}}
\put(780.00,341.00){\usebox{\plotpoint}}
\put(800,341){\usebox{\plotpoint}}
\put(780.00,364.00){\usebox{\plotpoint}}
\put(800,364){\usebox{\plotpoint}}
\put(821.00,333.00){\usebox{\plotpoint}}
\put(821,353){\usebox{\plotpoint}}
\put(811.00,333.00){\usebox{\plotpoint}}
\put(831,333){\usebox{\plotpoint}}
\put(811.00,353.00){\usebox{\plotpoint}}
\put(831,353){\usebox{\plotpoint}}
\end{picture}
}
\setbox2=\vbox{
\setlength{\unitlength}{0.240900pt}
\ifx\plotpoint\undefined\newsavebox{\plotpoint}\fi
\sbox{\plotpoint}{\rule[-0.200pt]{0.400pt}{0.400pt}}%
\begin{picture}(900,600)(0,0)
\font\gnuplot=cmr10 at 10pt
\gnuplot
\sbox{\plotpoint}{\rule[-0.200pt]{0.400pt}{0.400pt}}%
\put(220.0,281.0){\rule[-0.200pt]{148.394pt}{0.400pt}}
\put(220.0,176.0){\rule[-0.200pt]{4.818pt}{0.400pt}}
\put(198,176){\makebox(0,0)[r]{$-0.2$}}
\put(816.0,176.0){\rule[-0.200pt]{4.818pt}{0.400pt}}
\put(220.0,281.0){\rule[-0.200pt]{4.818pt}{0.400pt}}
\put(198,281){\makebox(0,0)[r]{$0$}}
\put(816.0,281.0){\rule[-0.200pt]{4.818pt}{0.400pt}}
\put(220.0,386.0){\rule[-0.200pt]{4.818pt}{0.400pt}}
\put(198,386){\makebox(0,0)[r]{$0.2$}}
\put(816.0,386.0){\rule[-0.200pt]{4.818pt}{0.400pt}}
\put(220.0,491.0){\rule[-0.200pt]{4.818pt}{0.400pt}}
\put(198,491){\makebox(0,0)[r]{$0.4$}}
\put(816.0,491.0){\rule[-0.200pt]{4.818pt}{0.400pt}}
\put(220.0,68.0){\rule[-0.200pt]{0.400pt}{4.818pt}}
\put(220,23){\makebox(0,0){$-1$ }}
\put(220.0,557.0){\rule[-0.200pt]{0.400pt}{4.818pt}}
\put(343.0,68.0){\rule[-0.200pt]{0.400pt}{4.818pt}}
\put(343,23){\makebox(0,0){$-0.6$ }}
\put(343.0,557.0){\rule[-0.200pt]{0.400pt}{4.818pt}}
\put(466.0,68.0){\rule[-0.200pt]{0.400pt}{4.818pt}}
\put(466,23){\makebox(0,0){$-0.2$ }}
\put(466.0,557.0){\rule[-0.200pt]{0.400pt}{4.818pt}}
\put(590.0,68.0){\rule[-0.200pt]{0.400pt}{4.818pt}}
\put(590,23){\makebox(0,0){$0.2$ }}
\put(590.0,557.0){\rule[-0.200pt]{0.400pt}{4.818pt}}
\put(713.0,68.0){\rule[-0.200pt]{0.400pt}{4.818pt}}
\put(713,23){\makebox(0,0){$0.6$ }}
\put(713.0,557.0){\rule[-0.200pt]{0.400pt}{4.818pt}}
\put(836.0,68.0){\rule[-0.200pt]{0.400pt}{4.818pt}}
\put(836,23){\makebox(0,0){$1$ }}
\put(836.0,557.0){\rule[-0.200pt]{0.400pt}{4.818pt}}
\put(220.0,68.0){\rule[-0.200pt]{148.394pt}{0.400pt}}
\put(836.0,68.0){\rule[-0.200pt]{0.400pt}{122.618pt}}
\put(220.0,577.0){\rule[-0.200pt]{148.394pt}{0.400pt}}
\put(45,322){\makebox(0,0){$\;F_0$}}
\put(220.0,68.0){\rule[-0.200pt]{0.400pt}{122.618pt}}
\put(706,512){\makebox(0,0)[r]{1546 MeV$/c$}}
\put(750,512){\raisebox{-.8pt}{\makebox(0,0){$\Diamond$}}}
\put(282,497){\raisebox{-.8pt}{\makebox(0,0){$\Diamond$}}}
\put(405,149){\raisebox{-.8pt}{\makebox(0,0){$\Diamond$}}}
\put(528,166){\raisebox{-.8pt}{\makebox(0,0){$\Diamond$}}}
\put(651,179){\raisebox{-.8pt}{\makebox(0,0){$\Diamond$}}}
\put(774,242){\raisebox{-.8pt}{\makebox(0,0){$\Diamond$}}}
\put(728.0,512.0){\rule[-0.200pt]{15.899pt}{0.400pt}}
\put(728.0,502.0){\rule[-0.200pt]{0.400pt}{4.818pt}}
\put(794.0,502.0){\rule[-0.200pt]{0.400pt}{4.818pt}}
\put(282.0,416.0){\rule[-0.200pt]{0.400pt}{38.785pt}}
\put(272.0,416.0){\rule[-0.200pt]{4.818pt}{0.400pt}}
\put(272.0,577.0){\rule[-0.200pt]{4.818pt}{0.400pt}}
\put(405.0,68.0){\rule[-0.200pt]{0.400pt}{39.026pt}}
\put(395.0,68.0){\rule[-0.200pt]{4.818pt}{0.400pt}}
\put(395.0,230.0){\rule[-0.200pt]{4.818pt}{0.400pt}}
\put(528.0,83.0){\rule[-0.200pt]{0.400pt}{39.748pt}}
\put(518.0,83.0){\rule[-0.200pt]{4.818pt}{0.400pt}}
\put(518.0,248.0){\rule[-0.200pt]{4.818pt}{0.400pt}}
\put(651.0,118.0){\rule[-0.200pt]{0.400pt}{29.390pt}}
\put(641.0,118.0){\rule[-0.200pt]{4.818pt}{0.400pt}}
\put(641.0,240.0){\rule[-0.200pt]{4.818pt}{0.400pt}}
\put(774.0,202.0){\rule[-0.200pt]{0.400pt}{19.272pt}}
\put(764.0,202.0){\rule[-0.200pt]{4.818pt}{0.400pt}}
\put(764.0,282.0){\rule[-0.200pt]{4.818pt}{0.400pt}}
\sbox{\plotpoint}{\rule[-0.500pt]{1.000pt}{1.000pt}}%
\put(706,467){\makebox(0,0)[r]{1695 MeV$/c$}}
\put(750,467){\circle*{18}}
\put(251,415){\circle*{18}}
\put(312,211){\circle*{18}}
\put(374,383){\circle*{18}}
\put(436,325){\circle*{18}}
\put(497,204){\circle*{18}}
\put(559,304){\circle*{18}}
\put(620,234){\circle*{18}}
\put(682,222){\circle*{18}}
\put(744,265){\circle*{18}}
\put(805,245){\circle*{18}}
\multiput(728,467)(20.756,0.000){4}{\usebox{\plotpoint}}
\put(794,467){\usebox{\plotpoint}}
\put(728.00,477.00){\usebox{\plotpoint}}
\put(728,457){\usebox{\plotpoint}}
\put(794.00,477.00){\usebox{\plotpoint}}
\put(794,457){\usebox{\plotpoint}}
\multiput(251,338)(0.000,20.756){8}{\usebox{\plotpoint}}
\put(251,493){\usebox{\plotpoint}}
\put(241.00,338.00){\usebox{\plotpoint}}
\put(261,338){\usebox{\plotpoint}}
\put(241.00,493.00){\usebox{\plotpoint}}
\put(261,493){\usebox{\plotpoint}}
\multiput(312,133)(0.000,20.756){8}{\usebox{\plotpoint}}
\put(312,288){\usebox{\plotpoint}}
\put(302.00,133.00){\usebox{\plotpoint}}
\put(322,133){\usebox{\plotpoint}}
\put(302.00,288.00){\usebox{\plotpoint}}
\put(322,288){\usebox{\plotpoint}}
\multiput(374,313)(0.000,20.756){7}{\usebox{\plotpoint}}
\put(374,453){\usebox{\plotpoint}}
\put(364.00,313.00){\usebox{\plotpoint}}
\put(384,313){\usebox{\plotpoint}}
\put(364.00,453.00){\usebox{\plotpoint}}
\put(384,453){\usebox{\plotpoint}}
\multiput(436,258)(0.000,20.756){7}{\usebox{\plotpoint}}
\put(436,391){\usebox{\plotpoint}}
\put(426.00,258.00){\usebox{\plotpoint}}
\put(446,258){\usebox{\plotpoint}}
\put(426.00,391.00){\usebox{\plotpoint}}
\put(446,391){\usebox{\plotpoint}}
\multiput(497,137)(0.000,20.756){7}{\usebox{\plotpoint}}
\put(497,271){\usebox{\plotpoint}}
\put(487.00,137.00){\usebox{\plotpoint}}
\put(507,137){\usebox{\plotpoint}}
\put(487.00,271.00){\usebox{\plotpoint}}
\put(507,271){\usebox{\plotpoint}}
\multiput(559,239)(0.000,20.756){7}{\usebox{\plotpoint}}
\put(559,369){\usebox{\plotpoint}}
\put(549.00,239.00){\usebox{\plotpoint}}
\put(569,239){\usebox{\plotpoint}}
\put(549.00,369.00){\usebox{\plotpoint}}
\put(569,369){\usebox{\plotpoint}}
\multiput(620,177)(0.000,20.756){6}{\usebox{\plotpoint}}
\put(620,292){\usebox{\plotpoint}}
\put(610.00,177.00){\usebox{\plotpoint}}
\put(630,177){\usebox{\plotpoint}}
\put(610.00,292.00){\usebox{\plotpoint}}
\put(630,292){\usebox{\plotpoint}}
\multiput(682,173)(0.000,20.756){5}{\usebox{\plotpoint}}
\put(682,271){\usebox{\plotpoint}}
\put(672.00,173.00){\usebox{\plotpoint}}
\put(692,173){\usebox{\plotpoint}}
\put(672.00,271.00){\usebox{\plotpoint}}
\put(692,271){\usebox{\plotpoint}}
\multiput(744,229)(0.000,20.756){4}{\usebox{\plotpoint}}
\put(744,302){\usebox{\plotpoint}}
\put(734.00,229.00){\usebox{\plotpoint}}
\put(754,229){\usebox{\plotpoint}}
\put(734.00,302.00){\usebox{\plotpoint}}
\put(754,302){\usebox{\plotpoint}}
\multiput(805,218)(0.000,20.756){3}{\usebox{\plotpoint}}
\put(805,273){\usebox{\plotpoint}}
\put(795.00,218.00){\usebox{\plotpoint}}
\put(815,218){\usebox{\plotpoint}}
\put(795.00,273.00){\usebox{\plotpoint}}
\put(815,273){\usebox{\plotpoint}}
\end{picture}
}
\hbox{\hspace{-0.5cm}\box1\hspace{-7cm}\box2}
\vspace{.5cm}
\setbox3=\vbox{
\setlength{\unitlength}{0.240900pt}
\ifx\plotpoint\undefined\newsavebox{\plotpoint}\fi
\sbox{\plotpoint}{\rule[-0.200pt]{0.400pt}{0.400pt}}%
\begin{picture}(900,600)(0,0)
\font\gnuplot=cmr10 at 10pt
\gnuplot
\sbox{\plotpoint}{\rule[-0.200pt]{0.400pt}{0.400pt}}%
\put(220.0,323.0){\rule[-0.200pt]{148.394pt}{0.400pt}}
\put(220.0,141.0){\rule[-0.200pt]{4.818pt}{0.400pt}}
\put(198,141){\makebox(0,0)[r]{$-1$}}
\put(816.0,141.0){\rule[-0.200pt]{4.818pt}{0.400pt}}
\put(220.0,213.0){\rule[-0.200pt]{4.818pt}{0.400pt}}
\put(198,213){\makebox(0,0)[r]{$-0.6$}}
\put(816.0,213.0){\rule[-0.200pt]{4.818pt}{0.400pt}}
\put(220.0,286.0){\rule[-0.200pt]{4.818pt}{0.400pt}}
\put(198,286){\makebox(0,0)[r]{$-0.2$}}
\put(816.0,286.0){\rule[-0.200pt]{4.818pt}{0.400pt}}
\put(220.0,359.0){\rule[-0.200pt]{4.818pt}{0.400pt}}
\put(198,359){\makebox(0,0)[r]{$0.2$}}
\put(816.0,359.0){\rule[-0.200pt]{4.818pt}{0.400pt}}
\put(220.0,432.0){\rule[-0.200pt]{4.818pt}{0.400pt}}
\put(198,432){\makebox(0,0)[r]{$0.6$}}
\put(816.0,432.0){\rule[-0.200pt]{4.818pt}{0.400pt}}
\put(220.0,504.0){\rule[-0.200pt]{4.818pt}{0.400pt}}
\put(198,504){\makebox(0,0)[r]{$1$}}
\put(816.0,504.0){\rule[-0.200pt]{4.818pt}{0.400pt}}
\put(220.0,68.0){\rule[-0.200pt]{0.400pt}{4.818pt}}
\put(220,23){\makebox(0,0){$-1$ }}
\put(220.0,557.0){\rule[-0.200pt]{0.400pt}{4.818pt}}
\put(343.0,68.0){\rule[-0.200pt]{0.400pt}{4.818pt}}
\put(343,23){\makebox(0,0){$-0.6$ }}
\put(343.0,557.0){\rule[-0.200pt]{0.400pt}{4.818pt}}
\put(466.0,68.0){\rule[-0.200pt]{0.400pt}{4.818pt}}
\put(466,23){\makebox(0,0){$-0.2$ }}
\put(466.0,557.0){\rule[-0.200pt]{0.400pt}{4.818pt}}
\put(590.0,68.0){\rule[-0.200pt]{0.400pt}{4.818pt}}
\put(590,23){\makebox(0,0){$0.2$ }}
\put(590.0,557.0){\rule[-0.200pt]{0.400pt}{4.818pt}}
\put(713.0,68.0){\rule[-0.200pt]{0.400pt}{4.818pt}}
\put(713,23){\makebox(0,0){$0.6$ }}
\put(713.0,557.0){\rule[-0.200pt]{0.400pt}{4.818pt}}
\put(836.0,68.0){\rule[-0.200pt]{0.400pt}{4.818pt}}
\put(836,23){\makebox(0,0){$1$ }}
\put(836.0,557.0){\rule[-0.200pt]{0.400pt}{4.818pt}}
\put(220.0,68.0){\rule[-0.200pt]{148.394pt}{0.400pt}}
\put(836.0,68.0){\rule[-0.200pt]{0.400pt}{122.618pt}}
\put(220.0,577.0){\rule[-0.200pt]{148.394pt}{0.400pt}}
\put(45,322){\makebox(0,0){$\; C_{xx}$}}
\put(220.0,68.0){\rule[-0.200pt]{0.400pt}{122.618pt}}
\put(282,509){\raisebox{-.8pt}{\makebox(0,0){$\Diamond$}}}
\put(405,393){\raisebox{-.8pt}{\makebox(0,0){$\Diamond$}}}
\put(528,463){\raisebox{-.8pt}{\makebox(0,0){$\Diamond$}}}
\put(651,406){\raisebox{-.8pt}{\makebox(0,0){$\Diamond$}}}
\put(774,258){\raisebox{-.8pt}{\makebox(0,0){$\Diamond$}}}
\put(282.0,444.0){\rule[-0.200pt]{0.400pt}{31.076pt}}
\put(272.0,444.0){\rule[-0.200pt]{4.818pt}{0.400pt}}
\put(272.0,573.0){\rule[-0.200pt]{4.818pt}{0.400pt}}
\put(405.0,329.0){\rule[-0.200pt]{0.400pt}{31.076pt}}
\put(395.0,329.0){\rule[-0.200pt]{4.818pt}{0.400pt}}
\put(395.0,458.0){\rule[-0.200pt]{4.818pt}{0.400pt}}
\put(528.0,397.0){\rule[-0.200pt]{0.400pt}{31.799pt}}
\put(518.0,397.0){\rule[-0.200pt]{4.818pt}{0.400pt}}
\put(518.0,529.0){\rule[-0.200pt]{4.818pt}{0.400pt}}
\put(651.0,357.0){\rule[-0.200pt]{0.400pt}{23.608pt}}
\put(641.0,357.0){\rule[-0.200pt]{4.818pt}{0.400pt}}
\put(641.0,455.0){\rule[-0.200pt]{4.818pt}{0.400pt}}
\put(774.0,226.0){\rule[-0.200pt]{0.400pt}{15.418pt}}
\put(764.0,226.0){\rule[-0.200pt]{4.818pt}{0.400pt}}
\put(764.0,290.0){\rule[-0.200pt]{4.818pt}{0.400pt}}
\sbox{\plotpoint}{\rule[-0.500pt]{1.000pt}{1.000pt}}%
\put(251,293){\circle*{18}}
\put(312,176){\circle*{18}}
\put(374,304){\circle*{18}}
\put(436,271){\circle*{18}}
\put(497,268){\circle*{18}}
\put(559,461){\circle*{18}}
\put(620,517){\circle*{18}}
\put(682,463){\circle*{18}}
\put(744,372){\circle*{18}}
\put(805,251){\circle*{18}}
\multiput(251,231)(0.000,20.756){6}{\usebox{\plotpoint}}
\put(251,354){\usebox{\plotpoint}}
\put(241.00,231.00){\usebox{\plotpoint}}
\put(261,231){\usebox{\plotpoint}}
\put(241.00,354.00){\usebox{\plotpoint}}
\put(261,354){\usebox{\plotpoint}}
\multiput(312,114)(0.000,20.756){6}{\usebox{\plotpoint}}
\put(312,238){\usebox{\plotpoint}}
\put(302.00,114.00){\usebox{\plotpoint}}
\put(322,114){\usebox{\plotpoint}}
\put(302.00,238.00){\usebox{\plotpoint}}
\put(322,238){\usebox{\plotpoint}}
\multiput(374,247)(0.000,20.756){6}{\usebox{\plotpoint}}
\put(374,360){\usebox{\plotpoint}}
\put(364.00,247.00){\usebox{\plotpoint}}
\put(384,247){\usebox{\plotpoint}}
\put(364.00,360.00){\usebox{\plotpoint}}
\put(384,360){\usebox{\plotpoint}}
\multiput(436,218)(0.000,20.756){6}{\usebox{\plotpoint}}
\put(436,324){\usebox{\plotpoint}}
\put(426.00,218.00){\usebox{\plotpoint}}
\put(446,218){\usebox{\plotpoint}}
\put(426.00,324.00){\usebox{\plotpoint}}
\put(446,324){\usebox{\plotpoint}}
\multiput(497,215)(0.000,20.756){6}{\usebox{\plotpoint}}
\put(497,322){\usebox{\plotpoint}}
\put(487.00,215.00){\usebox{\plotpoint}}
\put(507,215){\usebox{\plotpoint}}
\put(487.00,322.00){\usebox{\plotpoint}}
\put(507,322){\usebox{\plotpoint}}
\multiput(559,409)(0.000,20.756){6}{\usebox{\plotpoint}}
\put(559,513){\usebox{\plotpoint}}
\put(549.00,409.00){\usebox{\plotpoint}}
\put(569,409){\usebox{\plotpoint}}
\put(549.00,513.00){\usebox{\plotpoint}}
\put(569,513){\usebox{\plotpoint}}
\multiput(620,471)(0.000,20.756){5}{\usebox{\plotpoint}}
\put(620,564){\usebox{\plotpoint}}
\put(610.00,471.00){\usebox{\plotpoint}}
\put(630,471){\usebox{\plotpoint}}
\put(610.00,564.00){\usebox{\plotpoint}}
\put(630,564){\usebox{\plotpoint}}
\multiput(682,424)(0.000,20.756){4}{\usebox{\plotpoint}}
\put(682,502){\usebox{\plotpoint}}
\put(672.00,424.00){\usebox{\plotpoint}}
\put(692,424){\usebox{\plotpoint}}
\put(672.00,502.00){\usebox{\plotpoint}}
\put(692,502){\usebox{\plotpoint}}
\multiput(744,343)(0.000,20.756){3}{\usebox{\plotpoint}}
\put(744,401){\usebox{\plotpoint}}
\put(734.00,343.00){\usebox{\plotpoint}}
\put(754,343){\usebox{\plotpoint}}
\put(734.00,401.00){\usebox{\plotpoint}}
\put(754,401){\usebox{\plotpoint}}
\multiput(805,229)(0.000,20.756){3}{\usebox{\plotpoint}}
\put(805,273){\usebox{\plotpoint}}
\put(795.00,229.00){\usebox{\plotpoint}}
\put(815,229){\usebox{\plotpoint}}
\put(795.00,273.00){\usebox{\plotpoint}}
\put(815,273){\usebox{\plotpoint}}
\end{picture}
}
\setbox4=\vbox{
\setlength{\unitlength}{0.240900pt}
\ifx\plotpoint\undefined\newsavebox{\plotpoint}\fi
\sbox{\plotpoint}{\rule[-0.200pt]{0.400pt}{0.400pt}}%
\begin{picture}(900,600)(0,0)
\font\gnuplot=cmr10 at 10pt
\gnuplot
\sbox{\plotpoint}{\rule[-0.200pt]{0.400pt}{0.400pt}}%
\put(220.0,323.0){\rule[-0.200pt]{148.394pt}{0.400pt}}
\put(220.0,141.0){\rule[-0.200pt]{4.818pt}{0.400pt}}
\put(198,141){\makebox(0,0)[r]{$-1$}}
\put(816.0,141.0){\rule[-0.200pt]{4.818pt}{0.400pt}}
\put(220.0,213.0){\rule[-0.200pt]{4.818pt}{0.400pt}}
\put(198,213){\makebox(0,0)[r]{$-0.6$}}
\put(816.0,213.0){\rule[-0.200pt]{4.818pt}{0.400pt}}
\put(220.0,286.0){\rule[-0.200pt]{4.818pt}{0.400pt}}
\put(198,286){\makebox(0,0)[r]{$-0.2$}}
\put(816.0,286.0){\rule[-0.200pt]{4.818pt}{0.400pt}}
\put(220.0,359.0){\rule[-0.200pt]{4.818pt}{0.400pt}}
\put(198,359){\makebox(0,0)[r]{$0.2$}}
\put(816.0,359.0){\rule[-0.200pt]{4.818pt}{0.400pt}}
\put(220.0,432.0){\rule[-0.200pt]{4.818pt}{0.400pt}}
\put(198,432){\makebox(0,0)[r]{$0.6$}}
\put(816.0,432.0){\rule[-0.200pt]{4.818pt}{0.400pt}}
\put(220.0,504.0){\rule[-0.200pt]{4.818pt}{0.400pt}}
\put(198,504){\makebox(0,0)[r]{$1$}}
\put(816.0,504.0){\rule[-0.200pt]{4.818pt}{0.400pt}}
\put(220.0,68.0){\rule[-0.200pt]{0.400pt}{4.818pt}}
\put(220,23){\makebox(0,0){$-1$ }}
\put(220.0,557.0){\rule[-0.200pt]{0.400pt}{4.818pt}}
\put(343.0,68.0){\rule[-0.200pt]{0.400pt}{4.818pt}}
\put(343,23){\makebox(0,0){$-0.6$ }}
\put(343.0,557.0){\rule[-0.200pt]{0.400pt}{4.818pt}}
\put(466.0,68.0){\rule[-0.200pt]{0.400pt}{4.818pt}}
\put(466,23){\makebox(0,0){$-0.2$ }}
\put(466.0,557.0){\rule[-0.200pt]{0.400pt}{4.818pt}}
\put(590.0,68.0){\rule[-0.200pt]{0.400pt}{4.818pt}}
\put(590,23){\makebox(0,0){$0.2$ }}
\put(590.0,557.0){\rule[-0.200pt]{0.400pt}{4.818pt}}
\put(713.0,68.0){\rule[-0.200pt]{0.400pt}{4.818pt}}
\put(713,23){\makebox(0,0){$0.6$ }}
\put(713.0,557.0){\rule[-0.200pt]{0.400pt}{4.818pt}}
\put(836.0,68.0){\rule[-0.200pt]{0.400pt}{4.818pt}}
\put(836,23){\makebox(0,0){$1$ }}
\put(836.0,557.0){\rule[-0.200pt]{0.400pt}{4.818pt}}
\put(220.0,68.0){\rule[-0.200pt]{148.394pt}{0.400pt}}
\put(836.0,68.0){\rule[-0.200pt]{0.400pt}{122.618pt}}
\put(220.0,577.0){\rule[-0.200pt]{148.394pt}{0.400pt}}
\put(45,322){\makebox(0,0){$\; C_{yy}$}}
\put(220.0,68.0){\rule[-0.200pt]{0.400pt}{122.618pt}}
\put(282,388){\raisebox{-.8pt}{\makebox(0,0){$\Diamond$}}}
\put(405,508){\raisebox{-.8pt}{\makebox(0,0){$\Diamond$}}}
\put(528,572){\raisebox{-.8pt}{\makebox(0,0){$\Diamond$}}}
\put(651,523){\raisebox{-.8pt}{\makebox(0,0){$\Diamond$}}}
\put(774,452){\raisebox{-.8pt}{\makebox(0,0){$\Diamond$}}}
\put(282.0,324.0){\rule[-0.200pt]{0.400pt}{30.835pt}}
\put(272.0,324.0){\rule[-0.200pt]{4.818pt}{0.400pt}}
\put(272.0,452.0){\rule[-0.200pt]{4.818pt}{0.400pt}}
\put(405.0,444.0){\rule[-0.200pt]{0.400pt}{31.076pt}}
\put(395.0,444.0){\rule[-0.200pt]{4.818pt}{0.400pt}}
\put(395.0,573.0){\rule[-0.200pt]{4.818pt}{0.400pt}}
\put(528.0,506.0){\rule[-0.200pt]{0.400pt}{17.104pt}}
\put(518.0,506.0){\rule[-0.200pt]{4.818pt}{0.400pt}}
\put(518.0,577.0){\rule[-0.200pt]{4.818pt}{0.400pt}}
\put(651.0,474.0){\rule[-0.200pt]{0.400pt}{23.608pt}}
\put(641.0,474.0){\rule[-0.200pt]{4.818pt}{0.400pt}}
\put(641.0,572.0){\rule[-0.200pt]{4.818pt}{0.400pt}}
\put(774.0,420.0){\rule[-0.200pt]{0.400pt}{15.418pt}}
\put(764.0,420.0){\rule[-0.200pt]{4.818pt}{0.400pt}}
\put(764.0,484.0){\rule[-0.200pt]{4.818pt}{0.400pt}}
\sbox{\plotpoint}{\rule[-0.500pt]{1.000pt}{1.000pt}}%
\put(251,280){\circle*{18}}
\put(312,420){\circle*{18}}
\put(374,287){\circle*{18}}
\put(436,270){\circle*{18}}
\put(497,449){\circle*{18}}
\put(559,471){\circle*{18}}
\put(620,552){\circle*{18}}
\put(682,484){\circle*{18}}
\put(744,451){\circle*{18}}
\put(805,446){\circle*{18}}
\multiput(251,218)(0.000,20.756){6}{\usebox{\plotpoint}}
\put(251,342){\usebox{\plotpoint}}
\put(241.00,218.00){\usebox{\plotpoint}}
\put(261,218){\usebox{\plotpoint}}
\put(241.00,342.00){\usebox{\plotpoint}}
\put(261,342){\usebox{\plotpoint}}
\multiput(312,357)(0.000,20.756){7}{\usebox{\plotpoint}}
\put(312,482){\usebox{\plotpoint}}
\put(302.00,357.00){\usebox{\plotpoint}}
\put(322,357){\usebox{\plotpoint}}
\put(302.00,482.00){\usebox{\plotpoint}}
\put(322,482){\usebox{\plotpoint}}
\multiput(374,230)(0.000,20.756){6}{\usebox{\plotpoint}}
\put(374,343){\usebox{\plotpoint}}
\put(364.00,230.00){\usebox{\plotpoint}}
\put(384,230){\usebox{\plotpoint}}
\put(364.00,343.00){\usebox{\plotpoint}}
\put(384,343){\usebox{\plotpoint}}
\multiput(436,217)(0.000,20.756){6}{\usebox{\plotpoint}}
\put(436,323){\usebox{\plotpoint}}
\put(426.00,217.00){\usebox{\plotpoint}}
\put(446,217){\usebox{\plotpoint}}
\put(426.00,323.00){\usebox{\plotpoint}}
\put(446,323){\usebox{\plotpoint}}
\multiput(497,395)(0.000,20.756){6}{\usebox{\plotpoint}}
\put(497,503){\usebox{\plotpoint}}
\put(487.00,395.00){\usebox{\plotpoint}}
\put(507,395){\usebox{\plotpoint}}
\put(487.00,503.00){\usebox{\plotpoint}}
\put(507,503){\usebox{\plotpoint}}
\multiput(559,419)(0.000,20.756){6}{\usebox{\plotpoint}}
\put(559,523){\usebox{\plotpoint}}
\put(549.00,419.00){\usebox{\plotpoint}}
\put(569,419){\usebox{\plotpoint}}
\put(549.00,523.00){\usebox{\plotpoint}}
\put(569,523){\usebox{\plotpoint}}
\multiput(620,505)(0.000,20.756){4}{\usebox{\plotpoint}}
\put(620,577){\usebox{\plotpoint}}
\put(610.00,505.00){\usebox{\plotpoint}}
\put(630,505){\usebox{\plotpoint}}
\put(610.00,577.00){\usebox{\plotpoint}}
\put(630,577){\usebox{\plotpoint}}
\multiput(682,445)(0.000,20.756){4}{\usebox{\plotpoint}}
\put(682,523){\usebox{\plotpoint}}
\put(672.00,445.00){\usebox{\plotpoint}}
\put(692,445){\usebox{\plotpoint}}
\put(672.00,523.00){\usebox{\plotpoint}}
\put(692,523){\usebox{\plotpoint}}
\multiput(744,422)(0.000,20.756){3}{\usebox{\plotpoint}}
\put(744,480){\usebox{\plotpoint}}
\put(734.00,422.00){\usebox{\plotpoint}}
\put(754,422){\usebox{\plotpoint}}
\put(734.00,480.00){\usebox{\plotpoint}}
\put(754,480){\usebox{\plotpoint}}
\multiput(805,424)(0.000,20.756){3}{\usebox{\plotpoint}}
\put(805,468){\usebox{\plotpoint}}
\put(795.00,424.00){\usebox{\plotpoint}}
\put(815,424){\usebox{\plotpoint}}
\put(795.00,468.00){\usebox{\plotpoint}}
\put(815,468){\usebox{\plotpoint}}
\end{picture}
}
\hbox{\hspace{-0.5cm}\box3\hspace{-7cm}\box4}
\vspace{.5cm}
\setbox5=\vbox{
\setlength{\unitlength}{0.240900pt}
\ifx\plotpoint\undefined\newsavebox{\plotpoint}\fi
\sbox{\plotpoint}{\rule[-0.200pt]{0.400pt}{0.400pt}}%
\begin{picture}(900,600)(0,0)
\font\gnuplot=cmr10 at 10pt
\gnuplot
\sbox{\plotpoint}{\rule[-0.200pt]{0.400pt}{0.400pt}}%
\put(220.0,345.0){\rule[-0.200pt]{148.394pt}{0.400pt}}
\put(220.0,179.0){\rule[-0.200pt]{4.818pt}{0.400pt}}
\put(198,179){\makebox(0,0)[r]{$-1$}}
\put(816.0,179.0){\rule[-0.200pt]{4.818pt}{0.400pt}}
\put(220.0,246.0){\rule[-0.200pt]{4.818pt}{0.400pt}}
\put(198,246){\makebox(0,0)[r]{$-0.6$}}
\put(816.0,246.0){\rule[-0.200pt]{4.818pt}{0.400pt}}
\put(220.0,312.0){\rule[-0.200pt]{4.818pt}{0.400pt}}
\put(198,312){\makebox(0,0)[r]{$-0.2$}}
\put(816.0,312.0){\rule[-0.200pt]{4.818pt}{0.400pt}}
\put(220.0,378.0){\rule[-0.200pt]{4.818pt}{0.400pt}}
\put(198,378){\makebox(0,0)[r]{$0.2$}}
\put(816.0,378.0){\rule[-0.200pt]{4.818pt}{0.400pt}}
\put(220.0,444.0){\rule[-0.200pt]{4.818pt}{0.400pt}}
\put(198,444){\makebox(0,0)[r]{$0.6$}}
\put(816.0,444.0){\rule[-0.200pt]{4.818pt}{0.400pt}}
\put(220.0,511.0){\rule[-0.200pt]{4.818pt}{0.400pt}}
\put(198,511){\makebox(0,0)[r]{$1$}}
\put(816.0,511.0){\rule[-0.200pt]{4.818pt}{0.400pt}}
\put(220.0,113.0){\rule[-0.200pt]{0.400pt}{4.818pt}}
\put(220,68){\makebox(0,0){$-1$}}
\put(220.0,557.0){\rule[-0.200pt]{0.400pt}{4.818pt}}
\put(343.0,113.0){\rule[-0.200pt]{0.400pt}{4.818pt}}
\put(343,68){\makebox(0,0){$-0.6$}}
\put(343.0,557.0){\rule[-0.200pt]{0.400pt}{4.818pt}}
\put(466.0,113.0){\rule[-0.200pt]{0.400pt}{4.818pt}}
\put(466,68){\makebox(0,0){$-0.2$}}
\put(466.0,557.0){\rule[-0.200pt]{0.400pt}{4.818pt}}
\put(590.0,113.0){\rule[-0.200pt]{0.400pt}{4.818pt}}
\put(590,68){\makebox(0,0){$0.2$}}
\put(590.0,557.0){\rule[-0.200pt]{0.400pt}{4.818pt}}
\put(713.0,113.0){\rule[-0.200pt]{0.400pt}{4.818pt}}
\put(713,68){\makebox(0,0){$0.6$}}
\put(713.0,557.0){\rule[-0.200pt]{0.400pt}{4.818pt}}
\put(836.0,113.0){\rule[-0.200pt]{0.400pt}{4.818pt}}
\put(836,68){\makebox(0,0){$1$}}
\put(836.0,557.0){\rule[-0.200pt]{0.400pt}{4.818pt}}
\put(220.0,113.0){\rule[-0.200pt]{148.394pt}{0.400pt}}
\put(836.0,113.0){\rule[-0.200pt]{0.400pt}{111.778pt}}
\put(220.0,577.0){\rule[-0.200pt]{148.394pt}{0.400pt}}
\put(45,345){\makebox(0,0){$\; C_{zz}$}}
\put(528,-22){\makebox(0,0){$\cos\vartheta_{\rm cm}$ }}
\put(220.0,113.0){\rule[-0.200pt]{0.400pt}{111.778pt}}
\put(282,342){\raisebox{-.8pt}{\makebox(0,0){$\Diamond$}}}
\put(405,117){\raisebox{-.8pt}{\makebox(0,0){$\Diamond$}}}
\put(528,132){\raisebox{-.8pt}{\makebox(0,0){$\Diamond$}}}
\put(651,158){\raisebox{-.8pt}{\makebox(0,0){$\Diamond$}}}
\put(774,306){\raisebox{-.8pt}{\makebox(0,0){$\Diamond$}}}
\put(282.0,283.0){\rule[-0.200pt]{0.400pt}{28.185pt}}
\put(272.0,283.0){\rule[-0.200pt]{4.818pt}{0.400pt}}
\put(272.0,400.0){\rule[-0.200pt]{4.818pt}{0.400pt}}
\put(405.0,113.0){\rule[-0.200pt]{0.400pt}{15.177pt}}
\put(395.0,113.0){\rule[-0.200pt]{4.818pt}{0.400pt}}
\put(395.0,176.0){\rule[-0.200pt]{4.818pt}{0.400pt}}
\put(528.0,113.0){\rule[-0.200pt]{0.400pt}{19.272pt}}
\put(518.0,113.0){\rule[-0.200pt]{4.818pt}{0.400pt}}
\put(518.0,193.0){\rule[-0.200pt]{4.818pt}{0.400pt}}
\put(651.0,113.0){\rule[-0.200pt]{0.400pt}{21.440pt}}
\put(641.0,113.0){\rule[-0.200pt]{4.818pt}{0.400pt}}
\put(641.0,202.0){\rule[-0.200pt]{4.818pt}{0.400pt}}
\put(774.0,277.0){\rule[-0.200pt]{0.400pt}{14.213pt}}
\put(764.0,277.0){\rule[-0.200pt]{4.818pt}{0.400pt}}
\put(764.0,336.0){\rule[-0.200pt]{4.818pt}{0.400pt}}
\sbox{\plotpoint}{\rule[-0.500pt]{1.000pt}{1.000pt}}%
\put(251,338){\circle*{18}}
\put(312,313){\circle*{18}}
\put(374,292){\circle*{18}}
\put(436,233){\circle*{18}}
\put(497,247){\circle*{18}}
\put(559,217){\circle*{18}}
\put(620,152){\circle*{18}}
\put(682,124){\circle*{18}}
\put(744,231){\circle*{18}}
\put(805,313){\circle*{18}}
\multiput(251,281)(0.000,20.756){6}{\usebox{\plotpoint}}
\put(251,394){\usebox{\plotpoint}}
\put(241.00,281.00){\usebox{\plotpoint}}
\put(261,281){\usebox{\plotpoint}}
\put(241.00,394.00){\usebox{\plotpoint}}
\put(261,394){\usebox{\plotpoint}}
\multiput(312,256)(0.000,20.756){6}{\usebox{\plotpoint}}
\put(312,369){\usebox{\plotpoint}}
\put(302.00,256.00){\usebox{\plotpoint}}
\put(322,256){\usebox{\plotpoint}}
\put(302.00,369.00){\usebox{\plotpoint}}
\put(322,369){\usebox{\plotpoint}}
\multiput(374,241)(0.000,20.756){5}{\usebox{\plotpoint}}
\put(374,344){\usebox{\plotpoint}}
\put(364.00,241.00){\usebox{\plotpoint}}
\put(384,241){\usebox{\plotpoint}}
\put(364.00,344.00){\usebox{\plotpoint}}
\put(384,344){\usebox{\plotpoint}}
\multiput(436,185)(0.000,20.756){5}{\usebox{\plotpoint}}
\put(436,282){\usebox{\plotpoint}}
\put(426.00,185.00){\usebox{\plotpoint}}
\put(446,185){\usebox{\plotpoint}}
\put(426.00,282.00){\usebox{\plotpoint}}
\put(446,282){\usebox{\plotpoint}}
\multiput(497,198)(0.000,20.756){5}{\usebox{\plotpoint}}
\put(497,296){\usebox{\plotpoint}}
\put(487.00,198.00){\usebox{\plotpoint}}
\put(507,198){\usebox{\plotpoint}}
\put(487.00,296.00){\usebox{\plotpoint}}
\put(507,296){\usebox{\plotpoint}}
\multiput(559,170)(0.000,20.756){5}{\usebox{\plotpoint}}
\put(559,265){\usebox{\plotpoint}}
\put(549.00,170.00){\usebox{\plotpoint}}
\put(569,170){\usebox{\plotpoint}}
\put(549.00,265.00){\usebox{\plotpoint}}
\put(569,265){\usebox{\plotpoint}}
\multiput(620,113)(0.000,20.756){4}{\usebox{\plotpoint}}
\put(620,194){\usebox{\plotpoint}}
\put(610.00,113.00){\usebox{\plotpoint}}
\put(630,113){\usebox{\plotpoint}}
\put(610.00,194.00){\usebox{\plotpoint}}
\put(630,194){\usebox{\plotpoint}}
\multiput(682,113)(0.000,20.756){3}{\usebox{\plotpoint}}
\put(682,160){\usebox{\plotpoint}}
\put(672.00,113.00){\usebox{\plotpoint}}
\put(692,113){\usebox{\plotpoint}}
\put(672.00,160.00){\usebox{\plotpoint}}
\put(692,160){\usebox{\plotpoint}}
\multiput(744,205)(0.000,20.756){3}{\usebox{\plotpoint}}
\put(744,258){\usebox{\plotpoint}}
\put(734.00,205.00){\usebox{\plotpoint}}
\put(754,205){\usebox{\plotpoint}}
\put(734.00,258.00){\usebox{\plotpoint}}
\put(754,258){\usebox{\plotpoint}}
\multiput(805,293)(0.000,20.756){2}{\usebox{\plotpoint}}
\put(805,333){\usebox{\plotpoint}}
\put(795.00,293.00){\usebox{\plotpoint}}
\put(815,293){\usebox{\plotpoint}}
\put(795.00,333.00){\usebox{\plotpoint}}
\put(815,333){\usebox{\plotpoint}}
\end{picture}
}
\setbox6=\vbox{
\setlength{\unitlength}{0.240900pt}
\ifx\plotpoint\undefined\newsavebox{\plotpoint}\fi
\sbox{\plotpoint}{\rule[-0.200pt]{0.400pt}{0.400pt}}%
\begin{picture}(900,600)(0,0)
\font\gnuplot=cmr10 at 10pt
\gnuplot
\sbox{\plotpoint}{\rule[-0.200pt]{0.400pt}{0.400pt}}%
\put(220.0,345.0){\rule[-0.200pt]{148.394pt}{0.400pt}}
\put(220.0,179.0){\rule[-0.200pt]{4.818pt}{0.400pt}}
\put(198,179){\makebox(0,0)[r]{$-1$}}
\put(816.0,179.0){\rule[-0.200pt]{4.818pt}{0.400pt}}
\put(220.0,246.0){\rule[-0.200pt]{4.818pt}{0.400pt}}
\put(198,246){\makebox(0,0)[r]{$-0.6$}}
\put(816.0,246.0){\rule[-0.200pt]{4.818pt}{0.400pt}}
\put(220.0,312.0){\rule[-0.200pt]{4.818pt}{0.400pt}}
\put(198,312){\makebox(0,0)[r]{$-0.2$}}
\put(816.0,312.0){\rule[-0.200pt]{4.818pt}{0.400pt}}
\put(220.0,378.0){\rule[-0.200pt]{4.818pt}{0.400pt}}
\put(198,378){\makebox(0,0)[r]{$0.2$}}
\put(816.0,378.0){\rule[-0.200pt]{4.818pt}{0.400pt}}
\put(220.0,444.0){\rule[-0.200pt]{4.818pt}{0.400pt}}
\put(198,444){\makebox(0,0)[r]{$0.6$}}
\put(816.0,444.0){\rule[-0.200pt]{4.818pt}{0.400pt}}
\put(220.0,511.0){\rule[-0.200pt]{4.818pt}{0.400pt}}
\put(198,511){\makebox(0,0)[r]{$1$}}
\put(816.0,511.0){\rule[-0.200pt]{4.818pt}{0.400pt}}
\put(220.0,113.0){\rule[-0.200pt]{0.400pt}{4.818pt}}
\put(220,68){\makebox(0,0){$-1$}}
\put(220.0,557.0){\rule[-0.200pt]{0.400pt}{4.818pt}}
\put(343.0,113.0){\rule[-0.200pt]{0.400pt}{4.818pt}}
\put(343,68){\makebox(0,0){$-0.6$}}
\put(343.0,557.0){\rule[-0.200pt]{0.400pt}{4.818pt}}
\put(466.0,113.0){\rule[-0.200pt]{0.400pt}{4.818pt}}
\put(466,68){\makebox(0,0){$-0.2$}}
\put(466.0,557.0){\rule[-0.200pt]{0.400pt}{4.818pt}}
\put(590.0,113.0){\rule[-0.200pt]{0.400pt}{4.818pt}}
\put(590,68){\makebox(0,0){$0.2$}}
\put(590.0,557.0){\rule[-0.200pt]{0.400pt}{4.818pt}}
\put(713.0,113.0){\rule[-0.200pt]{0.400pt}{4.818pt}}
\put(713,68){\makebox(0,0){$0.6$}}
\put(713.0,557.0){\rule[-0.200pt]{0.400pt}{4.818pt}}
\put(836.0,113.0){\rule[-0.200pt]{0.400pt}{4.818pt}}
\put(836,68){\makebox(0,0){$1$}}
\put(836.0,557.0){\rule[-0.200pt]{0.400pt}{4.818pt}}
\put(220.0,113.0){\rule[-0.200pt]{148.394pt}{0.400pt}}
\put(836.0,113.0){\rule[-0.200pt]{0.400pt}{111.778pt}}
\put(220.0,577.0){\rule[-0.200pt]{148.394pt}{0.400pt}}
\put(45,345){\makebox(0,0){\shortstack{$\langle C_{xz},$\\$ C_{zx}  
\rangle$}}}
\put(528,-22){\makebox(0,0){$\cos\vartheta_{\rm cm}$ }}
\put(220.0,113.0){\rule[-0.200pt]{0.400pt}{111.778pt}}
\put(282,342){\raisebox{-.8pt}{\makebox(0,0){$\Diamond$}}}
\put(405,420){\raisebox{-.8pt}{\makebox(0,0){$\Diamond$}}}
\put(528,332){\raisebox{-.8pt}{\makebox(0,0){$\Diamond$}}}
\put(651,261){\raisebox{-.8pt}{\makebox(0,0){$\Diamond$}}}
\put(774,286){\raisebox{-.8pt}{\makebox(0,0){$\Diamond$}}}
\put(282.0,301.0){\rule[-0.200pt]{0.400pt}{19.995pt}}
\put(272.0,301.0){\rule[-0.200pt]{4.818pt}{0.400pt}}
\put(272.0,384.0){\rule[-0.200pt]{4.818pt}{0.400pt}}
\put(405.0,379.0){\rule[-0.200pt]{0.400pt}{19.995pt}}
\put(395.0,379.0){\rule[-0.200pt]{4.818pt}{0.400pt}}
\put(395.0,462.0){\rule[-0.200pt]{4.818pt}{0.400pt}}
\put(528.0,289.0){\rule[-0.200pt]{0.400pt}{20.476pt}}
\put(518.0,289.0){\rule[-0.200pt]{4.818pt}{0.400pt}}
\put(518.0,374.0){\rule[-0.200pt]{4.818pt}{0.400pt}}
\put(651.0,229.0){\rule[-0.200pt]{0.400pt}{15.177pt}}
\put(641.0,229.0){\rule[-0.200pt]{4.818pt}{0.400pt}}
\put(641.0,292.0){\rule[-0.200pt]{4.818pt}{0.400pt}}
\put(774.0,266.0){\rule[-0.200pt]{0.400pt}{9.877pt}}
\put(764.0,266.0){\rule[-0.200pt]{4.818pt}{0.400pt}}
\put(764.0,307.0){\rule[-0.200pt]{4.818pt}{0.400pt}}
\sbox{\plotpoint}{\rule[-0.500pt]{1.000pt}{1.000pt}}%
\put(251,335){\circle*{18}}
\put(312,337){\circle*{18}}
\put(374,378){\circle*{18}}
\put(436,287){\circle*{18}}
\put(497,405){\circle*{18}}
\put(559,400){\circle*{18}}
\put(620,389){\circle*{18}}
\put(682,275){\circle*{18}}
\put(744,273){\circle*{18}}
\put(805,347){\circle*{18}}
\multiput(251,295)(0.000,20.756){4}{\usebox{\plotpoint}}
\put(251,375){\usebox{\plotpoint}}
\put(241.00,295.00){\usebox{\plotpoint}}
\put(261,295){\usebox{\plotpoint}}
\put(241.00,375.00){\usebox{\plotpoint}}
\put(261,375){\usebox{\plotpoint}}
\multiput(312,297)(0.000,20.756){4}{\usebox{\plotpoint}}
\put(312,377){\usebox{\plotpoint}}
\put(302.00,297.00){\usebox{\plotpoint}}
\put(322,297){\usebox{\plotpoint}}
\put(302.00,377.00){\usebox{\plotpoint}}
\put(322,377){\usebox{\plotpoint}}
\multiput(374,342)(0.000,20.756){4}{\usebox{\plotpoint}}
\put(374,415){\usebox{\plotpoint}}
\put(364.00,342.00){\usebox{\plotpoint}}
\put(384,342){\usebox{\plotpoint}}
\put(364.00,415.00){\usebox{\plotpoint}}
\put(384,415){\usebox{\plotpoint}}
\multiput(436,252)(0.000,20.756){4}{\usebox{\plotpoint}}
\put(436,321){\usebox{\plotpoint}}
\put(426.00,252.00){\usebox{\plotpoint}}
\put(446,252){\usebox{\plotpoint}}
\put(426.00,321.00){\usebox{\plotpoint}}
\put(446,321){\usebox{\plotpoint}}
\multiput(497,370)(0.000,20.756){4}{\usebox{\plotpoint}}
\put(497,439){\usebox{\plotpoint}}
\put(487.00,370.00){\usebox{\plotpoint}}
\put(507,370){\usebox{\plotpoint}}
\put(487.00,439.00){\usebox{\plotpoint}}
\put(507,439){\usebox{\plotpoint}}
\multiput(559,366)(0.000,20.756){4}{\usebox{\plotpoint}}
\put(559,433){\usebox{\plotpoint}}
\put(549.00,366.00){\usebox{\plotpoint}}
\put(569,366){\usebox{\plotpoint}}
\put(549.00,433.00){\usebox{\plotpoint}}
\put(569,433){\usebox{\plotpoint}}
\multiput(620,359)(0.000,20.756){3}{\usebox{\plotpoint}}
\put(620,419){\usebox{\plotpoint}}
\put(610.00,359.00){\usebox{\plotpoint}}
\put(630,359){\usebox{\plotpoint}}
\put(610.00,419.00){\usebox{\plotpoint}}
\put(630,419){\usebox{\plotpoint}}
\multiput(682,250)(0.000,20.756){3}{\usebox{\plotpoint}}
\put(682,300){\usebox{\plotpoint}}
\put(672.00,250.00){\usebox{\plotpoint}}
\put(692,250){\usebox{\plotpoint}}
\put(672.00,300.00){\usebox{\plotpoint}}
\put(692,300){\usebox{\plotpoint}}
\multiput(744,254)(0.000,20.756){2}{\usebox{\plotpoint}}
\put(744,291){\usebox{\plotpoint}}
\put(734.00,254.00){\usebox{\plotpoint}}
\put(754,254){\usebox{\plotpoint}}
\put(734.00,291.00){\usebox{\plotpoint}}
\put(754,291){\usebox{\plotpoint}}
\multiput(805,333)(0.000,20.756){2}{\usebox{\plotpoint}}
\put(805,362){\usebox{\plotpoint}}
\put(795.00,333.00){\usebox{\plotpoint}}
\put(815,333){\usebox{\plotpoint}}
\put(795.00,362.00){\usebox{\plotpoint}}
\put(815,362){\usebox{\plotpoint}}
\end{picture}
}
\hbox{\hspace{-0.5cm}\box5\hspace{-7cm}\box6}
\vskip 1cm

%
%
\caption{\label{fig1}Polarisation $P$, spin-singlet fraction $F_0$,
  and spin-correlation coefficients $C_{ij}$ for the reaction
  $\bar{\rm p}{\rm p}\rightarrow \overline{\Lambda}\Lambda$ at 1446
  (open squares) and 1695 MeV$/c$ (black dots). Data are from Ref.\
  \protect\cite{Barnes91}.}
\end{figure}
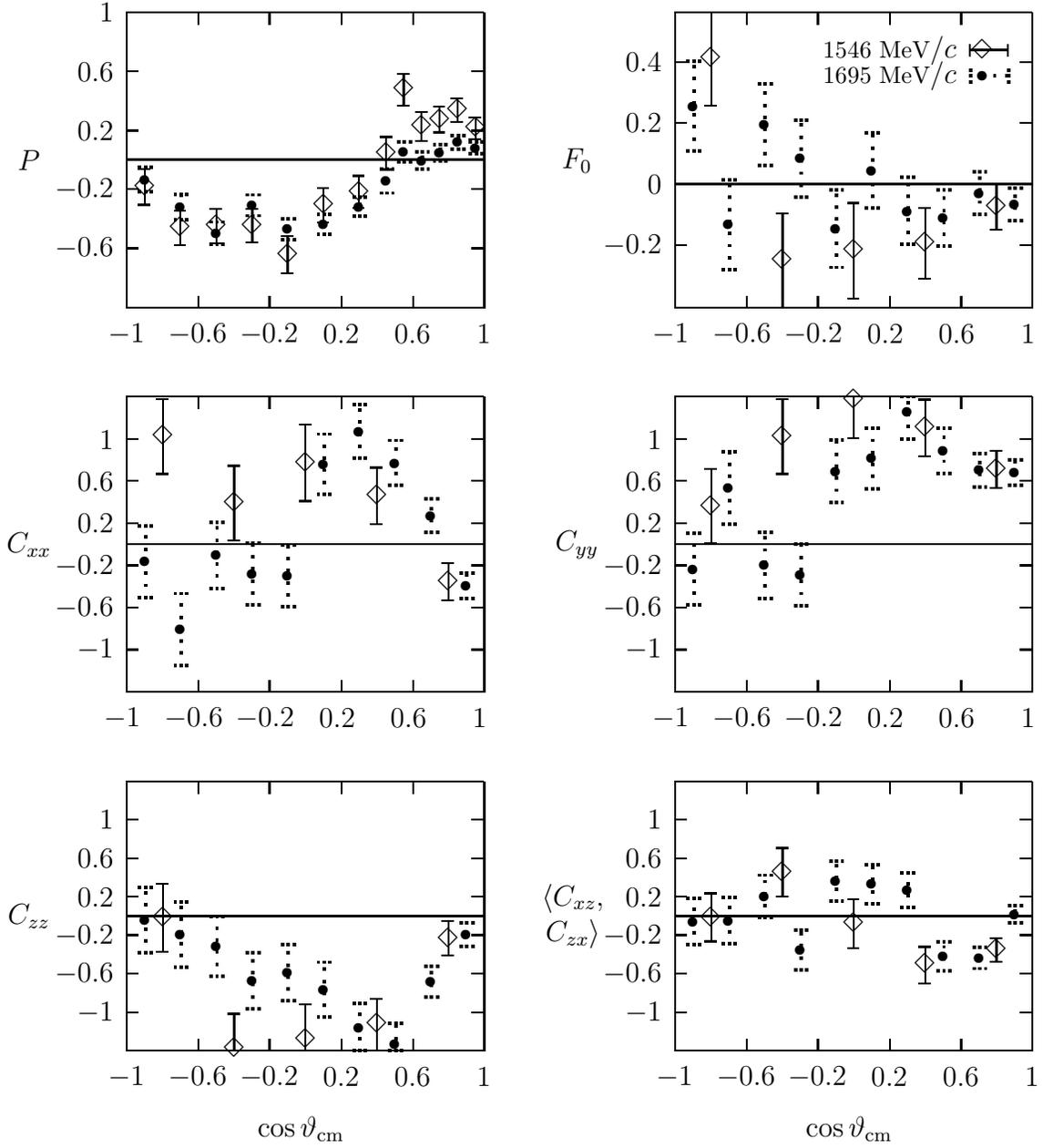

There is a renewed interest in the strangeness-exchange reaction
$\bar{\rm p}{\rm p} \rightarrow \overline{\Lambda}{\Lambda}$, due to
the results of PS185 collaboration at CERN, which has measured the
cross-section $\sigma$, the angular distribution $I_0(\vartheta_{\rm
  cm})$, polarisation $P(\vartheta_{\rm cm})$, and correlation
coefficients $C_{ij}(\vartheta_{\rm cm})$, at various energies
\cite{Fisher93}. Here $\vartheta_{\rm cm}$ is  the angle of
$\overline{\Lambda}$ with respect to $\bar{\rm p}$ in the
centre-of-mass frame.  Some of these data \cite{Barnes91} are  
reminded
in Fig.\ \ref{fig1}, at two selected momenta.  So far, the spin
measurements are given for free, thanks to the intrinsic asymmetry in
the decay of $\Lambda$ or $\overline{\Lambda}$. The experimental
investigation will be resumed using a transversely-polarised proton
target \cite{Bassalleck95}, giving access to new observables, in  
particular the
depolarisation $D_{nn}$ and spin-transfer $K_{nn}$ coefficients, not
to mention quantities involving more than two spins. The $D_{nn}$ and
$K_{nn}$ parameters measure by how much the transverse polarisation  
of
the proton target modifies the transverse polarisation of the
outcoming $\Lambda$ and $\overline{\Lambda}$, respectively.

As for the theory of $\bar{\rm p}{\rm p} \rightarrow
\overline{\Lambda}{\Lambda}$, there are many interesting
contributions, of which we shall only give a brief survey, before
focusing on those where depolarisation and spin-transfer are
discussed.  A  comprehensive bibliography can be found in Refs.\
\cite{Fisher93,Haidenbauer92a}. There are  two main approaches.
The first one relies on the exchange of strange mesons K, K$^*$, etc.
In the second one, the flavour-exchange process is described at the
quark level: a $\bar{\rm u}{\rm u}$ pair annihilates and a $\bar{\rm
  s}{\rm s}$ pair is created.  In both cases, one should account for
the strong absorption in the initial and final states.  Although  
these
two pictures are not necessarily conflicting, one would like to  
single
out the most efficient and the most realistic one. From the now
copious literature on the subject, one gets the impression that the
result of the competition is  a tie, as both Yukawa-type and
quark-inspired models claim to fit the PS185 data reasonably well.

In Ref.\ \cite{Haidenbauer92a}, Holinde et al.\ studied to which  
extent new
observables would help differentiate between the various models.
Among their conclusions, they insist on the role of $D_{nn}$ for
testing the underlying dynamics. Models with strange-meson exchange,
and models with $\bar{\rm s}{\rm s}$-pair creation give different
predictions. The former ones have $D_{nn}$ not far from its lower  
bound $-1$,
while the latter produce positive or slightly negative values, i.e.,  
a
less pronounced effect (remember that $D_{nn}=1$ in absence of
spin-dependent forces).

The analysis of Ref.\ \cite{Haidenbauer92a} might be understood as
follows. Current quark models impose constraints on the partial wave  
in
which the ${\rm u}\bar{\rm u}$ pair annihilates and the ${\rm
  s}\bar{\rm s}$ one is created. This is translated into correlations
between the outcoming hadrons, and presumably on the incoming
particles (in the event where both beam and target could be
polarised), but the spin wave-function tends to flow unchanged during
the reaction. On the other hand, the vertex linking a pseudoscalar
meson to a baryon line favours spin-flip transitions. In the
charge-exchange reaction $\bar{\rm p}{\rm p}\rightarrow\bar{\rm  
n}{\rm
  n}$, one expects large depolarisation or transfer coefficient,
especially in the longitudinal direction \cite{Richard81}. The data  
on
the transverse polarisation $D_{nn}$ of charge exchange already show
some interesting patterns \cite{Birsa93}.

The selectivity of $D_{nn}$ in $\bar{\rm p}{\rm p} \rightarrow
\overline{\Lambda}{\Lambda}$ was emphasised again recently by Alberg  
et al.\
 \cite{Alberg95}. They proposed a variant of the quark
model, where the $\bar{\rm s}{\rm s}$-pair is not created, but
extracted from the polarised sea of the incoming proton. They stated
that a depolarisation $D_{nn}=-1$ can be reached in an ``idealised
version'' of their picture. One could object that in half of the
cases, the $\bar{\rm s}{\rm s}$ pair comes from the antiproton,  
rather
than from the proton. However, in this model, as well as in the data,
the reaction takes place in a state of spin triplet, where the proton
and antiproton spins are aligned \cite{Alberg95a}.

An immediate and superficial observation is that a large negative
$D_{nn}$ would hardly separate the meson-exchange model of Holinde et
al.\ \cite{Haidenbauer92a} (and similar models in the literature)  
from
the polarised $\bar{\rm s}{\rm s}$-sea model of Alberg et al.\
\cite{Alberg95}.

The problem is  perhaps more serious. A closer look at Ref.\
\cite{Haidenbauer92a} shows that the meson-exchange model, though  
well
reproducing the smallness of the spin-singlet fraction
$F_0$, fails in accounting for the large negative values of the
correlation parameter $C_{zz}$. (This is confirmed, for instance, by
the results of LaFrance and Loiseau \cite{LaFrance91}, while a better
$C_{zz}$ is reached in the fit by Timmermans et al.\
\cite{Timmermans92}, who also use meson exchange.)\phantom{\@. }One  
of
the quark models of Holinde et al.\ gives a reasonable agreement for
$C_{zz}$, as does, at least at some angles, the model of Roberts
\cite{Roberts91}, who uses a different sign convention.

So far, Alberg et al.\ \cite{Alberg95} restricted themselves to a
qualitative investigation, and did not elaborate a model inspired by
their mechanism where all observables can be computed. One can  
already
say, however, that such a model, as well as any other detailed model,
would never produce $D_{nn}\simeq-1$ in kinematical regions where
$C_{zz}\simeq-1$. The reason is simple. Up to an overall factor of  
flux
and phase-space which can be omitted,
the observables of interest are given by  
\cite{Bystricky78,Haidenbauer92a}
\begin{eqnarray}
\label{Observables}
 I_0&=&|a|^2+|b|^2+|c|^2+|d|^2+|e|^2+|g|^2\nonumber\\
 PI_0&=&2{\rm Re}(ae^*)+2{\rm Im}(dg^*)\nonumber\\
 C_{xx}I_0&=&-2{\rm Re}(ad^*+bc^*)-2{\rm Im}(ge^*)\nonumber\\
 C_{yy}I_0&=&|a|^2-|b|^2-|c|^2+|d|^2+|e|^2+|g|^2\nonumber\\
 C_{zz}I_0&=&2{\rm Re}(ad^*-bc^*)+2{\rm Im}(ge^*)\nonumber\\
 C_{xz}I_0&=&-2{\rm Re}(ag^*)-2{\rm Im}(ed^*)\nonumber\\
 D_{nn}I_0&=&|a|^2+|b|^2-|c|^2-|d|^2+|e|^2-|g|^2\\
 D_{xx}I_0&=&-2{\rm Re}(ab^*+cd^*)\nonumber\\
 D_{zz}I_0&=&-2{\rm Re}(ab^*-cd^*)\nonumber\\
 D_{xz}I_0&=&2{\rm Re}(cg^*)+2{\rm Im}(be^*)\nonumber\\
 K_{nn}I_0&=&|a|^2-|b|^2+|c|^2-|d|^2+|e|^2-|g|^2\nonumber\\
 K_{xx}I_0&=&-2{\rm Re}(ac^*+bd^*)\nonumber\\
 K_{zz}I_0&=&-2{\rm Re}(ac^*-bd^*)\nonumber\\
 K_{xz}I_0&=&-2{\rm Re}(bg^*)+2{\rm Im}(ec^*).\nonumber
\end{eqnarray}
in terms of appropriate amplitudes $a,b,c,d,e$, and $g$. For more
details, see, for instance, Refs.\ \cite{Haidenbauer92a,Tabakin85}.  
The
spin-singlet fraction is given by
\begin{equation}
F_0={1\over4}(1+C_{xx}-C_{yy}+C_{zz})={1\over2}|b-c|^2,
\end{equation}
and would be $F_0=1/4$ in absence of spin-dependent interaction.

A look at Eqs.\ (\ref{Observables}) shows that $D_{nn}=-1$ would
require $a=b=e=0$, immediately implying $C_{zz}=0$. More precisely,  
one easily
gets from Eq.\ (\ref{Observables})
\begin{equation}\label{disk}
  C_{zz}^2+D_{nn}^2\le 1,
\end{equation}
so that a large $|C_{zz}|$ forces a small $D_{nn}$. Hence the
conclusion of Ref.\ \cite{Haidenbauer92a} that ``a measurement of the
depolarisation parameter $D_{nn}$ [\dots] can discriminate between
meson-exchange and simple constituent-quark models'', and the claim  
of
Ref.\ \cite{Alberg95} that ``the measurement of $D_{nn}$ [\dots] may
test dynamical mechanisms invoked to explain the proton-spin puzzle''
require that $|C_{zz}|$ is not too large.  The data in  
Fig.~\ref{fig1}
show that $C_{zz}$ is close to $-1$ at some angles and energies.  
There
are, however, uncertainties in these measurements, reflected for
instance in values exceeding the unitarity bound $|C_{zz}|\le 1$.  
Moreover,
recent data \cite{Roehrich95} indicate values of $|C_{zz}|$
smaller than the published values of Ref.\ \cite{Barnes91},  
especially above
$1.9\;$GeV/$c$.

A relation similar to (\ref{disk})  can be written for $C_{xx}$ and  
$D_{nn}$,
or
for $C_{xz}$ and $D_{nn}$. However, $|C_{xx}|$ and $|C_{xz}|$ are
on the average slightly smaller than $|C_{zz}|$, and thus less
constraining on the yet unknown $D_{nn}$.

Using a transversely-polarised target also gives access to $K_{nn}$.  
A
large effect seems excluded. For instance, $K_{nn}=-1$ would require
$a=c=e=0$, and thus $C_{zz}=0$, in conflict with present data.

We also note that the large values of $C_{yy}$ in the forward
hemisphere  suggest that $|b|\sim |c|\sim 0$ there, implying the
vanishing of $D_{xx}$, $D_{zz}$, $K_{xx}$, and $K_{zz}$. The  
situation
remains more open in the backward hemisphere. Hence looking at
$\overline{\Lambda}\Lambda$ production with a  
longitudinally-polarised
target would be worth doing.

Several further investigations remain to be done. In particular, one
could make more precise the allowed domain for the observables, in a
hypercube $[-1,+1]^n$ of $n$ selected spin parameters, which all run  
a
priori between $-1$ and $+1$, but are correlated. Years ago,
Cohen-Tannoudji and Messiah \cite{Cohen-Tannoudji64} pointed out
these correlations and redundancies, in the very same reaction  
$\bar{\rm p}{\rm
p}
\rightarrow \overline{\Lambda}{\Lambda}$. They showed that measuring
all spins in the final state and checking that the proton is
unpolarised would provide an estimate of the unknown polarisation of
the incoming antiproton beam. The algebraic relations between the  
spin
observables have been written down in the slightly simpler case of
elastic nucleon--nucleon scattering \cite{Delaney75}, and the
generalisation to $\bar{\rm p}{\rm
  p}\rightarrow\overline{\Lambda}\Lambda$ seems straightforward.

The proposed extension of the
PS185 experiment with a polarised target \cite{Bassalleck95} looks  
very
promising.  We have for instance pointed out that for the companion  
reactions
$\bar{\rm p}{\rm p}\rightarrow \bar{\rm p}{\rm p}$ and $\bar{\rm
  p}{\rm p}\rightarrow \bar{\rm n}{\rm n}$, the lack of detailed
spin measurements makes the phenomenological analysis rather delicate
\cite{Richard95}. The foreseen extension of the $\bar{\rm p}{\rm
  p}\rightarrow\overline{\Lambda}\Lambda$ experiment will
enable cross-checks of the internal consistency of the spin  
parameters
which are measured. The need for improvements is seen in Fig.\
\ref{fig1}, for some $C_{ij}$ exceed the allowed domain $[-1,+1]$,  
and
$F_0$ is sometimes negative. The polarisation $P$ will perhaps be
better determined through the analysing power $A_n$ ($P=A_n$ by
symmetry arguments), whose measurement becomes possible with some
transverse polarisation of the target. Equalities or inequalities
involving $K_{nn}$, $D_{nn}$ and the $C_{ij}$ can help reducing the
errors on each parameter, leading to more reliable physics
conclusions. However, at energies and angles where $|C_{zz}|$ turns  
out to be
large, $D_{nn}$
and $K_{nn}$ are necessarily small, and thus cannot provide further
tests of the models.  For this purpose, the respective merits of
longitudinal vs.\ transverse polarisation of the target should be
reconsidered.

\vskip .3cm {\bf Acknowledgements.} I would like to thank P.~Minnaert
and J.~Soffer for useful references on spin physics, M.~Alberg and
K.~Holinde for information about their own studies, K.~Kilian and
T.~Sefzick for providing me with the data and the experimental
projects, and W.~Roberts for comments on the manuscript. Constructive
comments by J.~Ellis, D.~Kharzeev and K.~R{\"o}hrich on an early
version of this letter are gratefully acknowledged. I benefitted from
the generous support of the Alexander von Humboldt foundation and the
hospitality of the Bonn University, making possible discussions with
several colleagues there, in particular E.~Klempt.

\clearpage


\end{document}